\documentclass[reprint,pre,aps,superscriptaddress,amsmath,longbibliography]{revtex4-2}
\usepackage{graphicx,dcolumn,multirow,amsfonts}
\usepackage[colorlinks,linkcolor=blue,citecolor=blue,anchorcolor=blue,urlcolor=blue]{hyperref}

\begin{document}

\title{Evolving Fractal Dimensions in Iterative Bicolored Percolation}

\author{Shuo Wei}
\thanks{These two authors contributed equally to this paper.}
\affiliation{Department of Modern Physics, University of Science and Technology of China, Hefei, Anhui 230026, China}

\author{Haoyu Liu}
\thanks{These two authors contributed equally to this paper.}
\affiliation{School of Mathematical Sciences, Peking University, Beijing 100871, China}

\author{Xin Sun}
\email{xinsun@bicmr.pku.edu.cn}
\affiliation{Beijing International Center for Mathematical Research, Peking University, Beijing 100871, China}

\author{Youjin Deng}
\email{yjdeng@ustc.edu.cn}
\affiliation{Department of Modern Physics, University of Science and Technology of China, Hefei, Anhui 230026, China}
\affiliation{Hefei National Research Center for Physical Sciences at the Microscale, University of Science and Technology of China, Hefei, Anhui 230026, China}
\affiliation{Hefei National Laboratory, University of Science and Technology of China, Hefei, Anhui 230088, China}

\author{Ming Li}
\email{lim@hfut.edu.cn}
\affiliation{School of Physics, Hefei University of Technology, Hefei, Anhui 230009, China}

\date{\today}

\begin{abstract}
Criticality is traditionally regarded as an unstable, fine-tuned fixed point of the renormalization group. We introduce an iterative bicolored percolation process in two dimensions and show that it can both preserve criticality and transform fractal dimensions. Starting from critical configurations, such as the O$(n)$ loop and fuzzy Potts models, successive coarse-graining generates a hierarchy of distinct yet critical generations. Using the conformal loop ensemble, we derive exact, generation-dependent fractal dimensions, which are quantitatively confirmed by large-scale Monte Carlo simulations. The evolutionary trajectory depends not only on the universality class of the initial state but also on whether it possesses a two-state critical structure, leading to different critical exponents starting from site and bond percolation. These results establish a general geometric mechanism for evolving fractal dimensions, in which scale invariance persists across generations.
\end{abstract}

\maketitle


\emph{Introduction}.--Criticality is a ubiquitous phenomenon observed across diverse domains, from physical systems~\cite{Kardar2007,Yeomans1992}, to biological networks~\cite{Valverde2015}, neural activities~\cite{Cocchi2017}, and even social sciences~\cite{Newman2010}. This prevalence has captivated researchers across distinct disciplines, as criticality often corresponds to optimal functioning, maximal responsiveness, and efficient information propagation~\cite{Cocchi2017,Beggs2003}.

The concept of criticality, rooted in statistical physics, describes a state poised between order and disorder, where scale-invariant correlations emerge and system-wide fluctuations exhibit power-law behavior. As a consequence, criticality is often manifested in geometric properties, and percolation theory provides a remarkably versatile framework~\cite{Stauffer1991}. Renormalization group (RG) theory examines the behavior of a system under successive coarse-graining transformations. In this framework, criticality corresponds to a fixed point--a state where the system remains invariant under RG transformations~\cite{Kardar2007}. The critical fixed point, however, is unstable. This instability means that the system parameters (e.g., temperature or external field) must be fine-tuned; small deviations cause the RG flow to move away into either ordered or disordered states.

\begin{figure}
\includegraphics[width=\columnwidth]{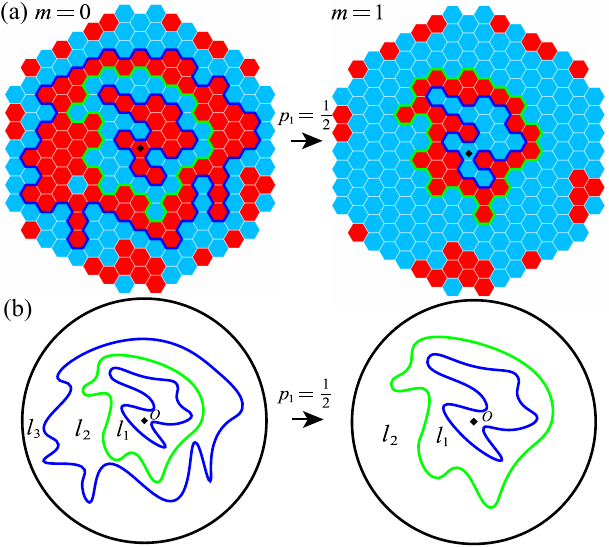}
\caption{(a) Schematic of the IBP process. Starting from a configuration with a two-state critical structure ($m=0$), where adjacent clusters have distinct states (colors), each cluster is independently recolored (with probability $p_1$), and neighboring clusters of the same color are merged to form the next generation ($m=1$). (b) In the loop representation, the IBP procedure corresponds to the stochastic elimination of loops (cluster boundaries). The loops surrounding the origin $O$ are the so-called nested loops.} \label{fig1}
\end{figure}

In this paper, we introduce an \textit{iterative bicolored percolation} (IBP) process in two dimensions (2D). We show that, starting from a critical initial configuration, the IBP process reorganizes the geometric structures of critical clusters, leading to continuously evolving fractal dimensions while preserving scale invariance. The process begins with a lattice where each site takes one of two possible states, denoted by red and blue for convenience (see Fig.~\ref{fig1} (a)). This initial assignment (generation $m=0$) is typically based on a specific physical process (e.g., site percolation or the Ising model). Adjacent sites of the same color form clusters, and adjacent clusters necessarily carry opposite colors. Subsequently, for each generation $m\geq 1$, we independently recolor each cluster from generation $(m-1)$ red with probability $p_m$ or blue with $1-p_m$. Adjacent clusters of the same color are then merged into larger ones. Note that the IBP process acts simultaneously across all length scales, rather than progressively eliminating short-distance degrees of freedom. This qualitatively distinguishes it from standard real-space renormalization-group coarse-graining procedures.

From a critical configuration, where the correlation length $\xi$ is of the order of the linear system size $L$ and the largest cluster spans the lattice, one might naturally expect the merging process to produce a giant cluster, leading to long-range order, given the known fragility of criticality. However, the numerical results in a recent work suggested the contrary~\cite{Li2024}. Starting with triangular-lattice site percolation at criticality, the critical geometric structures are not destroyed by IBP. In the thermodynamic limit ($L\to \infty$), criticality is preserved at any finite generation $m \geq 1$. The fractal dimension $d_f(m)$ of the percolation clusters increases monotonically, approaching $d_f(m\to \infty)=2$.

Generalizing the IBP process to any critical configurations, we establish that, for a broad class of 2D critical systems, the IBP process preserves criticality (scale invariance) while inducing a continuous evolution of fractal dimensions, highlighting a peculiar feature of planar graphs. We analyze two concrete examples: initial configurations corresponding to the O$(n)$ loop model~\cite{Nienhuis1982,Batchelor1989,Peled2019,DuminilCopin2021} and the fuzzy Potts model~\cite{Maes1995,Haeggstroem1999}. We derive the exact values of the generation-dependent fractal dimensions $d_f(m)$ using the conformal loop ensemble (CLE) approach~\cite{Sheffield2009,Nacu2011,Sheffield2012}. Extensive Monte Carlo simulations validate these theoretical predictions with excellent agreement. In contrast, when the $m=0$ configuration is non-critical, e.g., square-lattice site percolation with occupied probability $p=1/2 < p_c \approx 0.59$, we find that the system never attains criticality, even as the clusters grow larger.


\emph{Exact results for IBP of O$(n)$ loop model}.--Percolation models have long served as paradigms of critical geometry. At the percolation threshold, random scale-invariant structures emerge, providing realizations of criticality that connect to conformal field theory~\cite{Belavin1984,DiFrancesco1997}, Schramm-Loewner evolution (SLE)~\cite{Schramm2000}, and CLE~\cite{Sheffield2009,Sheffield2012}. In 2D, this geometric perspective has yielded a wealth of exact results, many established with full mathematical rigor.

Two prominent models closely related to percolation are the Fortuin-Kasteleyn (FK) random-cluster representation of the $Q$-state Potts model~\cite{Kasteleyn1969,Fortuin1972} and the loop formulation of the O$(n)$ spin model~\cite{Nienhuis1982,Batchelor1989,Peled2019,DuminilCopin2021}. These models are intimately related in 2D, sharing the same universality class when $Q=n^2$.

In the O$(n)$ loop model, typically defined on a honeycomb lattice, a bond is either occupied or empty. The constraint that each site has $0$ or $2$ incident bonds produces a gas of disjoint loops. The partition function is
\begin{equation}
{\cal Z}_{\rm loop} = \sum_{\rm loops} x^{{\cal N}_b} n^{{\cal N}_\ell} \; ,
\end{equation}
where ${\cal N}_b$ and ${\cal N}_\ell$ are the total numbers of occupied bonds and loops, respectively, and the sum runs over all loop configurations.

For $0 \le n \le 2$, increasing the bond weight $x$ drives a continuous phase transition at $x_+(n)$ from a dilute phase of small loops to a dense phase. In this dense phase ($x>x_+$), the system is critical, governed by the stable fixed point at $x_-(n)$. The free energy has been solved along the curves $1/x_{\pm} = \sqrt{2\pm \sqrt{2-n}}$~\cite{Nienhuis1982,Baxter1986,Baxter1987}. Along $x_+$ and $x_-$, the model corresponds to the \emph{tricritical} and \emph{critical} regimes of the Potts model with $Q=n^2$, respectively. In the Coulomb gas approach, the two regimes are described by the coupling strength $g$ (related to the SLE diffusion parameter $\kappa$ by $g\cdot \kappa=4$),
\begin{equation}
\sqrt{Q}=n= -2 \cos (\pi g)=-2 \cos (4\pi /\kappa) \; ,   \label{eq:Coulomb-gas-coupling}
\end{equation}
with $\kappa \in (8/3,4]$ for $x_{+}(n)$ and $\kappa \in [4,8)$ for $x_{-}(n)$.

The O$(n)$ loop model on the honeycomb lattice can also be interpreted as an extended Ising model on the dual triangular lattice~\cite{Baxter1986,Baxter1987,Fang2022}, where the loops correspond to the boundaries of Ising spin domains. In this picture, the red and blue clusters simply correspond to domains of the two Ising spin states. In particular, at $x_+(1)$ it is equivalent to the critical Ising model, while at $x_-(1)$ it reduces to critical site percolation.

In Fig.~\ref{fig1}(a), we sketch the O$(n)$ loop model on the honeycomb lattice. The simple (disjoint) loops naturally partition the lattice into clusters, which can then be colored in alternating red and blue so that the two sides of every loop carry different colors. Applying the IBP procedure to such a configuration, i.e., recoloring each cluster independently in red or blue, the neighboring clusters may acquire the same color and are then merged. In this merging step, the loop separating them is eliminated. Importantly, IBP never generates new loops; it only randomly removes existing ones, and the surviving loops preserve their geometric features.

We consider a region of radius $R$, as illustrated in Fig.~\ref{fig1}(a). Our focus is the one-arm exponent $\alpha_1$, which characterizes the power-law decay ($\sim R^{-\alpha_1}$) of the probability that the central site $O$ is connected to the boundary of the region by a path of sites of the same color as $R \to \infty$. The fractal dimension of red and blue clusters is then given by $d_f = 2 - \alpha_1$.

To calculate $\alpha_1$, we analyze the loops surrounding $O$, called nested loops (NL)~\cite{Nijs1983,Mitra2004,Ang2025,Song2025}. Let $\mathbb{P}_{0}(\ell)$ denote the probability of having $\ell$ NLs around $O$, with $\ell \ge 0$ an integer. Then, the one-arm exponent $\alpha_1$ is related to the probability of having no NLs,
\begin{equation}
\mathbb{P}_{0} (\ell=0) \sim R^{-\alpha_1}          \label{eq:p0a1}
\end{equation}

By introducing a real parameter $a$, one can define a continuous family of NL correlators,
\begin{equation}
{\cal W}_a \equiv \left\langle a^\ell \right\rangle \equiv \sum_{\ell \geq 0} a^\ell \, \mathbb{P}_{0} (\ell) \sim R^{-X_{\rm NL}(a)} \; ,
\label{eq:nested-loop-operator}
\end{equation}
where $X_{\rm NL}(a)$ is the $a$-dependent NL exponent. By definition, it has
\begin{equation}
X_{\rm NL}(0) = \alpha_1,
\end{equation}
and the probability normalization, $\sum_\ell \mathbb{P}_0(\ell) = 1$, enforces $X_{\rm NL}(1) = 0$. As explained in~\cite[Section 1.2]{Ang2025}
based on the CLE approach, the exact value of $X_{\rm NL}(a)$ for $a>0$ is the unique solution in the interval $\left(0, 1 - \frac{2}{\kappa} - \frac{3 \kappa}{32}\right)$ of
\begin{equation}
\cos \left(\pi \sqrt{\left(1-\frac{4}{\kappa}\right)^2+ \frac{8 X_{\rm NL}}{\kappa} } \right)
+a \cos \left( \frac{4 \pi}{\kappa} \right) = 0 \; .
\label{eq:one-arm-1}
\end{equation}

We now apply IBP to the red and blue domains on the triangular lattice. Similar to Eq.~(\ref{eq:p0a1}), the one-arm exponent for the $m$th generation is defined via
\begin{equation}
\mathbb{P}_m(0) \sim R^{-\alpha_1(m)},
\end{equation}
where $\mathbb{P}_m(0)$ is the probability that the number of NLs surrounding the origin is $\ell = 0$ at generation $m$. Since the IBP acts as a stochastic elimination of loops (see Fig.~\ref{fig1}), $\mathbb{P}_m(0)$ can be expressed in terms of the initial configuration ($m=0$) through the NL correlator summation
\begin{equation}
\mathbb{P}_m(0) = \sum_{\ell \ge 0} {\cal P}_{m,\ell} \, \mathbb{P}_0(\ell),
\label{eq:nested-loop-m}
\end{equation}
where $\mathbb{P}_0(\ell)$ is the probability of having $\ell$ NLs at $m=0$, and ${\cal P}_{m,\ell}$ is the probability that all $\ell$ NLs have been eliminated by generation $m$.

Once ${\cal P}_{m,\ell}$ is known, the parameter $a$ in Eq.~\eqref{eq:nested-loop-operator} can be determined by matching Eq.~\eqref{eq:nested-loop-m}. Substituting the resulting parameter for generation $m$, denoted by $a_m$, into Eq.~\eqref{eq:one-arm-1}, the one-arm exponent for generation $m$ is obtained as
\begin{equation}
\alpha_1(m) = X_{\rm NL}(a_m).
\end{equation}
We now present the explicit forms of ${\cal P}_{m,\ell}$, $a_m$, and the resulting exponents for two illustrative cases.

\emph{Symmetric case} ($p_m = 1/2$ for all $m$). In this case, each loop is independently preserved with probability $1/2$ at each generation. Consequently, a NL is eliminated after $m$ generations with probability $1-2^{-m}$, leading to
\begin{equation}
{\cal P}_{m,\ell} = \left(1 - \frac{1}{2^m}\right)^\ell.
\end{equation}
As a result, $\mathbb{P}_m(0)$ reduces exactly to the NL correlator ${\cal W}_{a_m}$, with $a_m=1-2^{-m}$. Then, we have the one-arm exponent $\alpha_1(m) = X_{\rm NL}(a_m)$.

\emph{Asymmetric case} ($1/2<p_m<1$). In this case, a closed-form expression of $\alpha_1(m)$ at arbitrary $m$ is not available. Here we present explicit results for the first two generations, $m=1$ and $2$.

For $m=1$, the $\ell$ NLs surrounding a site are eliminated with probability
\begin{equation}
{\cal P}_{1,\ell} = p_1^{\ell+1} + (1-p_1)^{\ell+1},
\end{equation}
which leads to
\begin{equation}
\mathbb{P}_1(0) = p_1 {\cal W}_{p_1} + (1-p_1) {\cal W}_{1-p_1}.
\end{equation}
Since $X_{\rm NL}(a)$ is a decreasing function of $a$, the contribution of ${\cal W}_{1-p_1}$ is asymptotically negligible, yielding the one-arm exponent $\alpha_1(1)=X_{\rm NL}(a_1)$ with $a_1=p_1$.

For $m=2$, a similar but more involved calculation gives $a_2=\frac{1+\sqrt{1-4p_1(1-p_1)(1-p_2^2)}}{2}$. The one-arm exponent for $m=2$ is thus given by $\alpha_1(2) = X_{\rm NL}(a_2)$.


\emph{Exact results for IBP of the fuzzy Potts model}.--The $Q$-state Potts model~\cite{Wu1982} is a paradigmatic system in statistical physics. Its Hamiltonian (reduced by the Boltzmann constant $k_{\rm B}$ and temperature $T$) is ${\cal H}/k_{\rm B}T = - K \sum_{\langle ij \rangle} \delta_{\sigma_i, \sigma_j}$, where $K>0$ is the ferromagnetic coupling, the spin $\sigma$ takes one of $Q$ possible states, $\delta$ is the Kronecker delta function, and the sum runs over all nearest-neighbor pairs. Using the FK representation, the Potts model can be reformulated as a geometric random-cluster model~\cite{Grimmett2006,DuminilCopin2021a}. The FK partition function reads
\begin{equation}
{\cal Z}_{\rm FK} = \sum_{\{\mathcal{G}\}} v^{{\cal N}_b} Q^{{\cal N}_k} \; ,
\label{eq:FK-partition}
\end{equation}
where the sum runs over all subsets $\mathcal{G}$ of occupied bonds, $v = e^K - 1$ is the bond weight, and ${\cal N}_b$ and ${\cal N}_k$ denote the total numbers of occupied bonds and connected components (FK clusters), respectively. The parameter $Q$ can be generalized to any real value $Q \in [0,\infty)$, and in the limit $Q \to 1$ the model reduces to standard bond percolation. In 2D, the FK-Potts model exhibits a continuous phase transition for $Q \in [0,4]$. At criticality, the scaling limit of FK cluster boundaries is described by a CLE with $\kappa \in [4,8)$, related to $Q$ via Eq.~\eqref{eq:Coulomb-gas-coupling}.

Unlike the O$(n)$ loop model, where loops naturally separate alternating-color domains, FK clusters have no intrinsic two-state structure. To apply IBP, we therefore first assign each FK cluster a color (red or blue), after which adjacent clusters with the same color merge. The resulting bicolored configuration is analogous to that of the O$(n)$ loop model and is taken as generation $m=0$ of the IBP process. By regarding the original FK configuration as generation $m=-1$, this initial coloring step can be uniformly viewed as an IBP update from $m=-1$ to $m=0$, where each cluster is independently colored red with probability $p_{0}$ and blue with probability $1-p_{0}$.

Although the $Q$-state Potts model is related to the O$(n)$ loop model, it does not allow one to derive the one-arm exponents starting from generation $m=-1$ using the same formulas as in the O$(n)$ case. The reason is that the IBP update from $m=-1$ to $m=0$ fundamentally changes the continuum description: from $\mathrm{CLE}_{\kappa}$ to variants of $\mathrm{CLE}_{\kappa' = 16/\kappa}$~\cite{KSL25,Miller2017}. In fact, generation $m=0$ is precisely the fuzzy Potts model~\cite{Maes1995,Haeggstroem1999} (also known as the divide-and-color model~\cite{Haeggstroem2001}). This duality $\kappa' = 16/\kappa$ reflects the fundamental difference between the FK boundaries at $m=-1$, which have self-intersections in the continuum limit, and the loops at $m=0$, which are simple (non-self-intersecting). Importantly, the initial coloring probability $p_{0}$ serves as an additional relevant parameter in the scaling limit of the fuzzy Potts model.

Specifically, we consider the symmetric initial coloring, $p_{0} = 1/2$. For this case, the exponent $X_{\rm NL}(a)$ for the fuzzy Potts model ($m=0$) can be deduced from~\cite[Section 2.2]{Liu2024} and is given by the unique positive solution in the interval $\left(0, 1 - \frac{\kappa'}{8} - \frac{3}{2\kappa'}\right)$ to a transcendental equation. Introducing
\begin{equation}
\theta = \frac{\pi}{\kappa'} \sqrt{(4-\kappa')^2 + 8 \kappa' X_{\rm NL}},
\end{equation}
the equation reads (see Eq.~(4.8) of~\cite{Cai2025}),
\begin{align}
&a \left[ \sin \left(\frac{\kappa'-1}{2} \theta\right)
- \sin \left(\frac{\kappa'-5}{2} \pi\right) \, \sin \left(\frac{\theta}{2}\right) \right] \nonumber \\
=& \sin \left(\frac{\kappa'+1}{2} \theta\right)
+ \sin \left(\frac{\kappa'-5}{2} \pi\right) \, \sin\left(\frac{\theta}{2}\right).
\label{eq:one-arm-FK}
\end{align}
For subsequent symmetric IBP iterations ($p_m = 1/2$ for all $m > 0$), an analogous argument to that following Eq.~\eqref{eq:nested-loop-m} shows that the one-arm exponent for any generation $m$ is given by $\alpha_1(m) = X_{\rm NL}(a_m)$, where $a_m=1-2^{-m}$.

For $Q=2$ ($\kappa=16/3$), generation $m=0$ corresponds to the critical Ising model, and Eq.~\eqref{eq:one-arm-FK} reduces to
\begin{equation}
2 \cos \left( \frac{\pi \sqrt{1 + 24 X_{\rm NL}}}{3} \right) = 1 - \frac{1}{2^m}.
\end{equation}
This result is consistent with the solution obtained from the O$(1)$ loop model by substituting $\kappa = 3$ into Eq.~\eqref{eq:one-arm-1}, confirming the known physical correspondence between the critical Ising model and the O$(1)$ loop model at the $x_{+}$ branch (tricritical $Q=1$ Potts model).


\begin{table}
\caption{Fitted fractal dimensions $d_f(m)$ for the IBP process with $p_m = 1/2$ for all generations $m$. The data without error bars represent the exact values $d_f(m) = 2 - \alpha_1(m)$.}   \label{tab:table1}
\begin{ruledtabular}
\begin{tabular}{clllll}
 &  \multicolumn{1}{c}{$m=-1$} & \multicolumn{1}{c}{$m=0$} & \multicolumn{1}{c}{$m=1$} & \multicolumn{1}{c}{$m=2$} & \multicolumn{1}{c}{$m=3$}     \\
\hline
\multicolumn{1}{c}{$n$}  &   \multicolumn{5}{c}{O$(n)$ loop model ($x_-$ branch)}     \\
$1$         &  & 1.89584(5)    & 1.9513(1)   & 1.9764(1)   & 1.9884(1)     \\
            &  & 1.895833...   & 1.95130...  & 1.97637...  & 1.98835...    \\
$\sqrt{2}$  &  & 1.8751(1)     & 1.9429(3)   & 1.97256(6)  & 1.98652(3)   \\
            &  & 1.875         & 1.94286...  & 1.972514... & 1.986504...  \\
$2$         &  & 1.8750(1)     & 1.9442(2)   & 1.9735(1)   & 1.9870(1)     \\
            &  & 1.875         & 1.94444...  & 1.97354...  & 1.98706...    \\
\hline
\multicolumn{1}{c}{$Q$}  &   \multicolumn{5}{c}{Fuzzy Potts model}  \\
$1$    & 1.89583(5)     & 1.9554(1)      & 1.97901(5)     & 1.98980(4)  &  1.99497(3)   \\
       & 1.895833...    & 1.95544...     & 1.979022...    & 1.989794... &  1.994962...  \\
$2$    & 1.8747(9)      & 1.9478(3)      & 1.9757(1)      & 1.9882(1)   &  1.9942(2)     \\
       & 1.875          & 1.94791...     & 1.97565...     & 1.98818...  &  1.99417...  \\
$3$    & 1.867(1)       & 1.944(1)       & 1.974(1)       & 1.987(1)    &  1.9937(5)    \\
       & 1.8666...      & 1.9445...      & 1.9740...      & 1.9874...   &  1.99379...
\end{tabular}
\end{ruledtabular}
\end{table}

\emph{Numerical experiments}.--Both the O$(n)$ loop model ($n \ge 1$) and the FK-Potts model ($Q \ge 1$) were simulated using cluster Monte Carlo algorithms. For the O$(n)$ loop model on the honeycomb lattice, simulations were implemented via the extended Ising model on triangular lattices~\cite{Fang2022}, and the resulting Ising domains were directly used as the IBP initial configuration ($m=0$). Critical slowing down is minor in this scheme and completely absent along the $x_{-}$ branch. For the $Q$-state Potts model on square lattices, the cluster algorithms~\cite{Swendsen1987,Wolff1989} suffer from significant critical slowing down for $Q \ge 3$, and finite-size corrections decay slowly with increasing system size; therefore, we restricted simulations to $Q \in [1,3]$. All simulations employed periodic boundary conditions on lattices up to linear size $L_{\max}=1024$.


\begin{figure}
\includegraphics[width=\columnwidth]{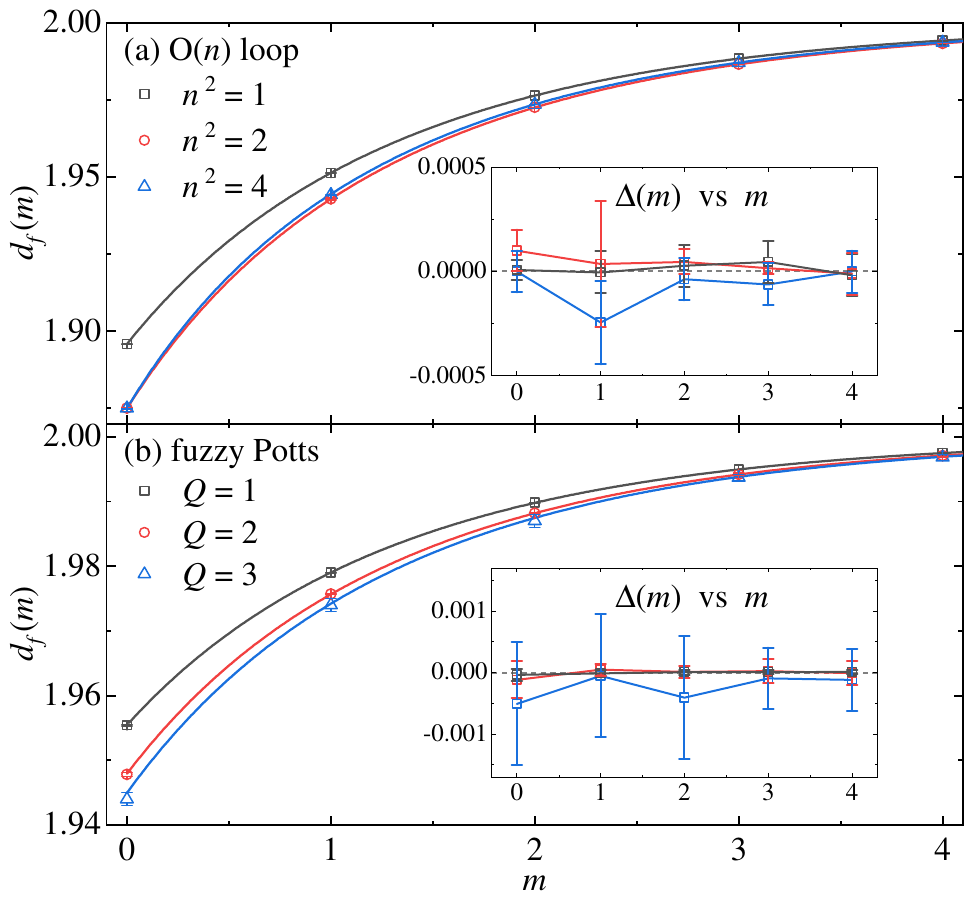}
\caption{Evolution of fractal dimensions $d_f(m)$ under the IBP process with $p_m=1/2$ for all generations $m$. (a) Results starting from O$(n)$ loop configurations at $x_{-}(n)$. (b) Results from fuzzy Potts configurations. Curves represent theoretical predictions. Insets: deviation between fitted and exact values of fractal dimension, $\Delta (m)=d_f^{\,\rm fit}(m)-d_f^{\,\rm th.}(m)$.} \label{fig2}
\end{figure}

Because precisely measuring the elimination probability $\mathbb{P}_m(0)$ is computationally costly, we determine the fractal dimensions $d_f(m) = 2 - \alpha_1(m)$ numerically via finite-size scaling of the largest cluster, $C_1(m) \sim L^{d_f(m)}$. The data are fitted to the scaling ansatz:
\begin{equation}
C_1(m) = L^{d_f(m)} (a_0+a_1L^{-\omega}),
\end{equation}
where $\omega$ is a correction exponent. Fits were performed using data with $L \ge L_{\min}$, and $L_{\min}$ was chosen as the smallest size for which the fit quality is reasonable and further increases do not significantly reduce $\chi^2$ per degree of freedom. The resulting estimates of $d_f(m)$ for the symmetric IBP are listed in Table~\ref{tab:table1}.

Figure~\ref{fig2} shows the fitted fractal dimensions $d_f(m)$ as a function of generation $m$. The numerical results fall accurately onto the theoretical curves, revealing distinct families of $d_f(m)$ for different initial configurations. The insets display the deviation between numerical values and theory, which is of order $10^{-4}$ for the O$(n)$ loop model and the fuzzy Potts model with $Q=1,2$, and of order $10^{-3}$ for $Q=3$, indicating excellent quantitative agreement. These deviations are much smaller than the differences of $d_f(m)$ between successive generations (of order $10^{-2}$), confirming the generation dependence of $d_f(m)$.

\begin{figure}
\includegraphics[width=\columnwidth]{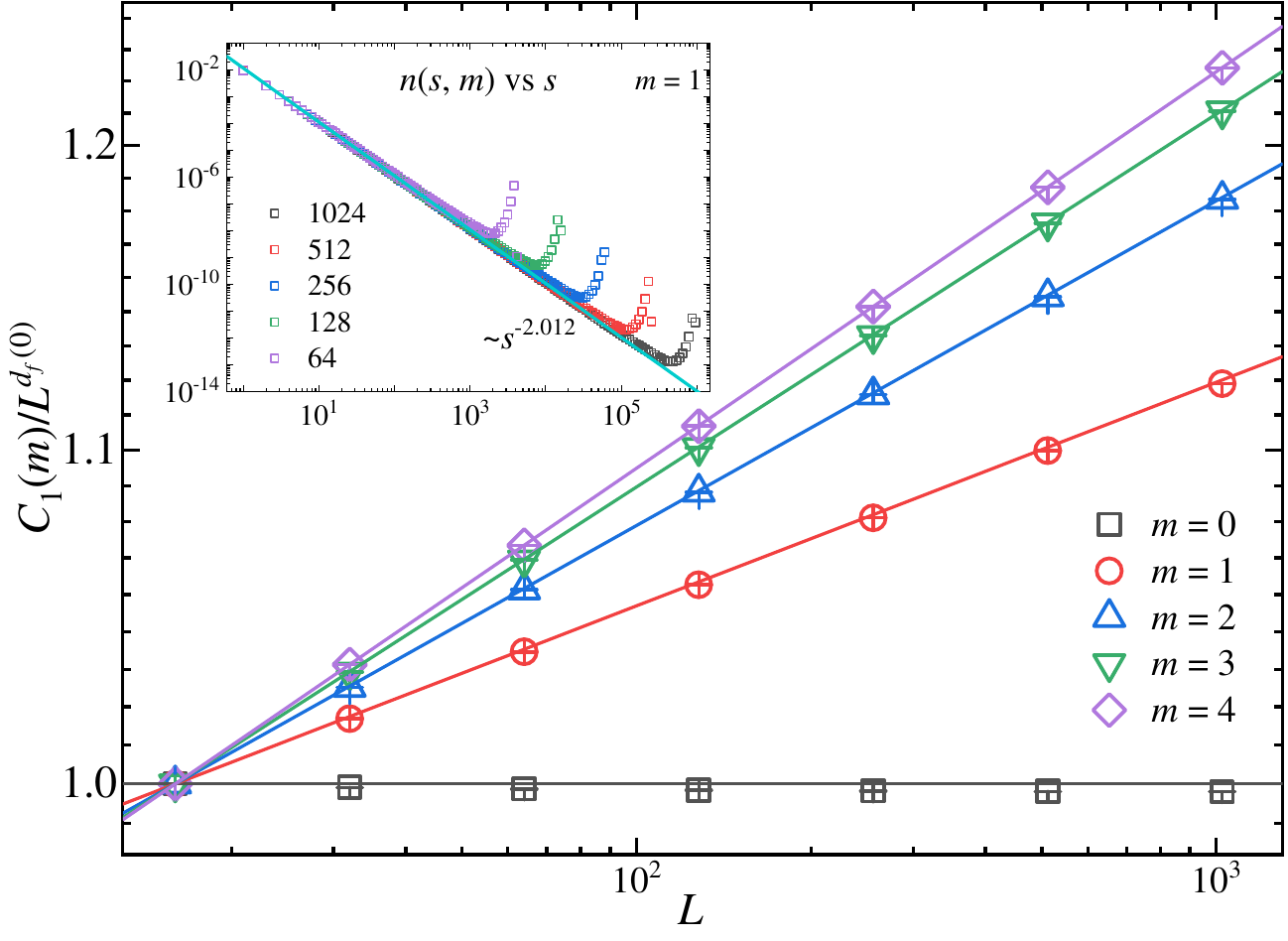}
\caption{Scaled size of the largest cluster, $C_1(m)/L^{d_f(0)}$, across generations $m$ for various system sizes $L$, starting from the critical Ising configuration (fuzzy Potts under $Q=2$). Lines represent theoretical predictions, $\sim L^{d_f(m)-d_f(0)}$. Data are normalized so that the first point of each generation equals unity. Inset: cluster number density $n(s,m)$ at $m=1$ for different $L$, defined as the number of clusters of size $s$ per site. The solid line indicates the scaling $\sim s^{-\tau(m)}$, with $\tau(m=1)\approx 2.012$. Using the relation $\tau=1+d/d_f$, this value is consistent with the theoretical fractal dimension $d_f(m=1)\approx1.97565$.}   \label{fig3}
\end{figure}

To further verify the generation-dependent exponents numerically, we plot the rescaled largest cluster size, $C_1(m)/L^{d_f(0)}$, versus system size $L$. As shown in Fig.~\ref{fig3} for the critical Ising model ($Q=2$), a clear scaling $\sim L^{d_f(m)-d_f(0)}$ is observed, in agreement with the theoretical predictions for $d_f(m)$.

In addition to the largest cluster, the criticality of each generation is confirmed by the power-law distribution of cluster sizes $n(s,m) \sim s^{-\tau(m)}$ (see, e.g., the inset of Fig.~\ref{fig3}), where the Fisher exponent is consistent with the fractal dimension via $\tau(m) = 1 + 2/d_f(m)$.

\begin{figure}
\includegraphics[width=\columnwidth]{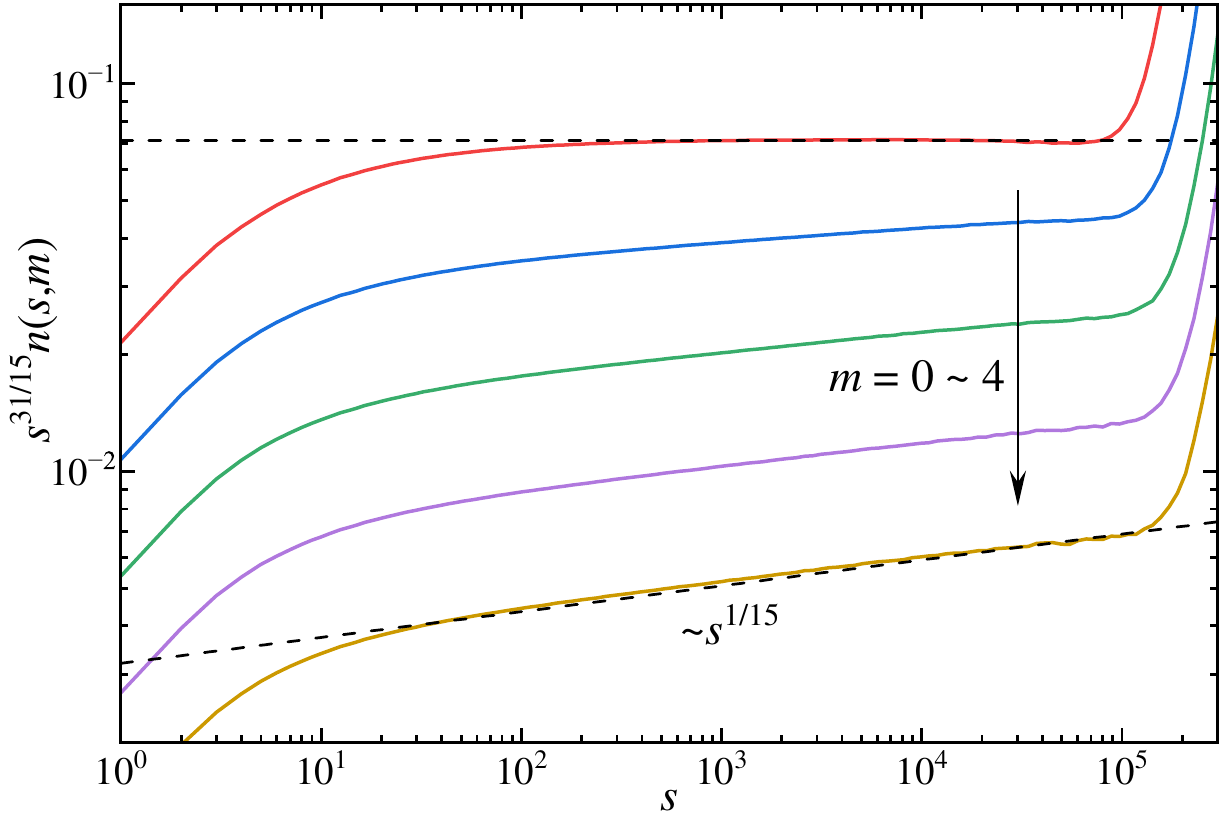}
\caption{The rescaled cluster-number density $s^{31/15} n(s,m)$ for different generations. Generation $m=0$ is the O($\sqrt{2}$) loop model at the $x_-$ branch, where $n(s,0)\sim s^{-31/15}$ (2D FK-Ising universality class). Thus, $s^{31/15} n(s,0)$ demonstrates a horizontal curve indicated by the top dashed line. Deviations at small and large $s$ are due to finite-size effects ($L=1024$). With increasing $m$, the curves become increasingly tilted, signaling a different Fisher exponent. The bottom dashed line ($\sim s^{1/15}$) indicates the limit case $n(s,\infty)\sim s^{-2}$.} \label{fig4}
\end{figure}

To visualize the evolution of the Fisher exponent, we further plot $s^{31/15} n(s,m)$ as a function of $s$ for different generations in Fig.~\ref{fig4}. For $m=0$, the system lies in the 2D FK-Ising universality class (O($\sqrt{2}$) loop model at the $x_-$ branch), where $n(s,m)\sim s^{-31/15}$, so that $s^{31/15} n(s,m)$ appears as a horizontal curve. With increasing $m$, the curves become increasingly tilted, signaling a continuously varying Fisher exponent. This indicates that each generation constitutes a genuine critical state, where scale invariance persists across all clusters, rather than merely featuring a single fractal largest cluster.

Moreover, the ability of IBP to maintain criticality across generations can also be understood through the perspective of self-matching. A detailed argument, along with extensive numerical evidence, can be found in Ref.~\cite{Li2024}, which uses critical site percolation on the triangular lattice as an example.

\emph{Discussion}.--The IBP process uncovers an unconventional aspect of criticality: it provides an exactly solvable setting in which a system remains conformally invariant while its critical exponents evolve continuously under simple stochastic dynamics. This stands in sharp contrast to the traditional view of isolated critical points, and establishes IBP as a stochastic coarse-graining procedure that generates an RG-like flow within the critical manifold. This flow is controlled mainly by the critical boundary geometry of the initial configuration. As a result, models that share the same universality class, like critical site and bond percolation, may nevertheless evolve along different IBP trajectories due to their distinct boundary structures. Note that the IBP process acts simultaneously across all length scales, distinguishing it qualitatively from standard real-space RG procedure.

Our study provides a framework for viewing the evolution of critical geometries and raises a number of questions whose resolution would deepen our understanding of criticality. First, it is not hard to argue using SLE/CLE techniques that the polychromatic arm exponents of IBP stay unchanged through the iteration. On the other hand, we expect the monochromatic arm exponents to be generation-dependent. The monochromatic two-arm exponent (i.e., the backbone exponent) of the Bernoulli percolation was recently derived in~\cite{Nolin2025}. We expect that the method can be extended to the IBP case. Understanding monochromatic exponents with more arms remains open. Second, the critical behavior of 2D random loop and cluster models is governed by conformal field theories, whose rich structure is still under intensive investigation~\cite{Jacobsen2025}. We expect the IBP will further generate an interesting class of conformal field theories. In particular, we observe that the IBP process with rational parameters $p_m$ typically produces a transcendental exponent $\alpha_1(m)$ for $m\ge 1$.


More broadly, IBP suggests an extended notion of universality: rather than isolated fixed points, entire families of critical states may be connected by dynamical flows, hinting at a new paradigm for the geometry and evolution of critical systems.


\emph{Acknowledgment}.--Y.D., S.W., and M.L. have been supported by the National Natural Science Foundation of China (under Grant No.\ 12275263), the Innovation Program for Quantum Science and Technology (under Grant No.\ 2021ZD0301900), the Natural Science Foundation of Fujian Province of China (under Grant No.\ 2023J02032). X.S. and H.L.\ have been supported by National Key R\&D Program of China (No.\ 2023YFA1010700).

\bibliography{ref}

\end{document}